\documentclass[twocolumn,showpacs,showkeys,amsmath,amssymb,superscriptaddress,floatfix,nofootinbib]{revtex4}

\usepackage{bm}
\usepackage{amsmath}
\usepackage{graphicx}
\usepackage{subfigure}
\usepackage[usenames,dvipsnames]{color}
\usepackage[colorlinks=true,linktocpage=true,linkcolor=blue,citecolor=blue]{hyperref}

\usepackage{setspace}
\usepackage{footmisc}
\usepackage[makeroom]{cancel}
\usepackage{comment}

\usepackage[normalem]{ulem} 
\definecolor{aoenglish}{rgb}{0.0, 0.5, 0.0}

\newcommand{\ped}{{\cal E}}
\newcommand{\pres}{{\cal P}}
\newcommand{\pedeq}{\ped_{\rm eq}}
\newcommand{\preseq}{\pres_{\rm eq}}
\newcommand{\thetapar}{\theta_\|}
\newcommand{\enlpar}{\vphantom{\frac{}{}}}

\newcommand{\taueq}{\tau_{\rm eq}}

\newcommand{\vp}{\enlpar}

\begin{document}

\title[Non-boost-invariant dissipative hydrodynamics]{ 
Non-boost-invariant dissipative hydrodynamics}

\author{Wojciech Florkowski}
\author{Radoslaw Ryblewski}
\affiliation{Institute of Nuclear Physics, Polish Academy of Sciences, PL-31342 Krak\'ow, Poland}

\author{Michael Strickland} 
\affiliation{Department of Physics, Kent State University, Kent, OH 44242 United States}

\author{Leonardo Tinti} 
\affiliation{Institute of Physics, Jan Kochanowski University, PL-25406~Kielce, Poland}

\begin{abstract}

The one-dimensional non-boost-invariant evolution of the quark-gluon plasma, presumably produced during the early stages of heavy-ion collisions, is analyzed within the frameworks of viscous and anisotropic hydrodynamics. We neglect transverse dynamics and assume homogeneous conditions in the transverse plane but, differently from Bjorken expansion, we relax longitudinal boost invariance in order to study the rapidity dependence of various hydrodynamical observables. We compare the results obtained using several formulations of second-order viscous hydrodynamics with a recent approach to anisotropic hydrodynamics, which treats the large initial pressure anisotropy in a non-perturbative fashion.  The results obtained with second-order viscous hydrodynamics depend on the particular choice of the second-order terms included, which suggests that the latter should be included in the most complete way. The results of anisotropic hydrodynamics and viscous hydrodynamics agree for the central hot part of the system, however, they differ at the edges where the approach of anisotropic hydrodynamics helps to control the undesirable growth of viscous corrections observed in standard frameworks.

\end{abstract}


\maketitle  

\section{Introduction}

 The ultimate goal of the ultra-relativistic heavy-ion collision experiments at the Relativistic Heavy Ion Collider (RHIC) and the Large Hadron Collider (LHC)  is to investigate the properties of nuclear  matter at extreme conditions of very high temperature and energy density. The collective behavior observed in these experiments has been described, at first, using perfect-fluid relativistic hydrodynamics.  Later, second-order viscous hydrodynamics was used based on general arguments that the shear viscosity to entropy density ratio is bounded from below \cite{Danielewicz:1984ww}.  The resulting second-order viscous hydrodynamics codes resulted in better agreement between hydrodynamic predictions and the experimental data. 

In contrast to the basic assumption of small deviations from perfect-fluid hydrodynamics which is used in the derivation of viscous hydrodynamic equations, the rapid longitudinal expansion, typically found in relativistic heavy-ion collisions, is the source of large pressure anisotropies. To address such large pressure corrections more accurately, a new framework called anisotropic hydrodynamics has been developed~\cite{Florkowski:2010cf,Martinez:2010sc}. The predictions of anisotropic hydrodynamics have been compared with the exact solutions of the Boltzmann equation for longitudinally boost invariant and transversely homogeneous systems. It has been shown that anisotropic hydrodynamics describes the dynamics more accurately than standard viscous frameworks~\cite{Florkowski:2013lza,Florkowski:2013lya}. Very similar results have been found  also for the Gubser type of expansion~\cite{Nopoush:2014qba,Denicol:2014tha,Denicol:2014xca}. Also recently, it has been shown that anisotropic hydrodynamics shows better agreement with the data for a completely different physical system, namely, an ultra-cold Fermi gas~\cite{Schaefer:2014awa,Bluhm:2015raa,Bluhm:2015bzi,Schaefer:2016yzd}.

Longitudinal boost invariance seems to be a reasonable approximation for the midrapidity region during the initial stages of collisions~\cite{Bjorken:1982qr,Baym:1984sr}, but it must be broken at large rapidities, because of the finite size of the expanding fireball (finite available energy). The gradients in the rapidity direction provide a new source of pressure corrections that affect the hydrodynamical evolution. In this paper, we investigate how different formulations of viscous hydrodynamics and anisotropic hydrodynamics deal with such corrections. We consider the M\"uller-Israel-Stewart (MIS) formulation of the second-order viscous hydrodynamics~\cite{Muller:1967zza,Israel:1976tn,Israel:1979wp} that is most common in phenomenological studies~\cite{Muronga:2001zk,Heinz:2005bw,
Bozek:2009dw,Schenke:2011bn,Karpenko:2013wva,Shen:2014vra}, as well as two alternative approaches, namely, the modified 14-moment expansion developed by Denicol et al.~\cite{Denicol:2010xn,Denicol:2012cn,Denicol:2014loa} (DNMR) and the gradient expansion presented in~\cite{Baier:2007ix} (BRSSS). For the anisotropic hydrodynamics framework, we choose the latest construction of the leading-order terms, which is based on the anisotropic matching principle~\cite{Tinti:2015xwa}. Below, we shall refer to this framework as AHYDRO. For sake of simplicity, we consider all formulations in the conformal limit, with the kinetic coefficients obtained from the kinetic theory in the relaxation time approximation (RTA)~\cite{Bhatnagar:1954zz} with classical statistics. 

The non-boost-invariant one-dimensional expansion of matter produced in heavy-ion collisions has been already studied for perfect-fluid~\cite{Florkowski:1992yd,Satarov:2006iw}, viscous~\cite{Bozek:2007qt}, as well as anisotropic hydrodynamics \cite{Ryblewski:2010bs,Martinez:2010sd}, see also~\cite{Gorenstein:1977xv}. In Ref.~\cite{Florkowski:1992yd} the convective stability of matter was analysed in the case with finite baryon number density. In Ref.~\cite{Bozek:2007qt} the emphasis was placed on the reduction of the longitudinal pressure due to the viscous effects as compared to perfect-fluid case \cite{Satarov:2006iw}, with consequences for the estimates of the initial energy density of the system created in heavy-ion collisions. 

In Refs.~\cite{Ryblewski:2010bs,Martinez:2010sd} the first formulations of anisotropic hydrodynamics for non-boost-invariant systems were introduced. Compared to these works, the AHYDRO framework used in this work is based on the more recent approach~\cite{Tinti:2015xwa}, that agrees with DNMR and BRSSS for systems approaching local thermal equilibrium. In fact, it has been recently demonstrated that the gradient expansions of DNMR, BRSSS and AHYDRO agree up to the second order in gradients for boost-invariant systems~\cite{Florkowski:2016zsi}. More recently, the perturbations of the baryon number density around the Bjorken solution were studied in \cite{Floerchinger:2015efa}.

In this work we investigate the generation of negative longitudinal pressure at large space-time rapidity $\eta$ which, in all second-order viscous hydrodynamics considered, results in matter ``in-flow'' and the development of a shock front.  AHYDRO eliminates negative pressures by construction and delays substantially the formation of a shock (to times much larger than the typical time span of heavy-ion collisions).  The negative pressures are physically not allowed in the kinetic calculations used to extract the values of our transport coefficients, however, they are present at the edges of the expanding system if standard viscous hydrodynamics is used.  We find that MIS and BRSSS results are very similar and lead to the largest negative pressures --- a change from positive to negative pressure has the character of a shock, where pressure as well as the fluid rapidity change very suddenly in a narrow range of $\eta$. The DNMR prescription works better, leading to smaller negative values of the pressure. The overall better performance of DNMR may be related to the fact that the DNMR equations are directly derived from the underlying RTA kinetic theory. It should be noted also that different treatment of the shear-to-shear coupling terms in the evolution of the shear stress tensor produces, contrary to common assumptions, a significant difference in the evolution of the system, especially for the flow profile at the edges of the system. 

AHYDRO regulates the dynamics of the expanding system. The system's pressure at large space-time rapidities is positive (although very close to zero), while the fluid rapidity is continuous and very close to $\eta$ at the system's edges. This allows for a continuous and consistent description of the produced system as consisting of the hot central part together with a free-streaming halo. 

The paper is organized as follows: In Sec.~\ref{sect:genprinc} the general principles of relativistic hydrodynamics based on the conservation laws are introduced. The constraints coming from a simple (1+1)-dimensional expansion geometry are implemented in Sec.~\ref{sect:implement}. Different versions of second-order viscous hydrodynamics are introduced and discussed in Sec.~\ref{sect:secondorder}, while anisotropic hydrodynamics is presented in Sec.~\ref{sect:ahydro}. Our numerical results are shown and discussed in Sec.~\ref{sect:results}. We summarize and conclude in Sec.~\ref{sect:conclusions}. Throughout the paper we use natural units with $c=\hbar=k_B=1$ and the metric tensor is $g_{\mu\nu} = \hbox{diag}(1,-1,-1,-1)$.
%
\section{General principles}
\label{sect:genprinc}
%
In the present work we neglect conserved charges (such as baryon number), hence, the main hydrodynamic equations reflect the conservation of  energy and momentum in the system, 
\begin{equation}
 \partial_\mu T^{\mu\nu}(x) = 0,
 \label{en_mom}
\end{equation}
where $T^{\mu\nu}$ is the energy-momentum tensor. We further use the Landau definition of the four-velocity~$U^\mu$,
\begin{equation}
 U_\mu(x) T^{\mu\nu}(x) =  \ped (x)U^\nu(x),
 \label{Landau_frame}
\end{equation}
where the eigenvalue $\ped$ is the proper energy density. Then, we make use of the most general decomposition of $T^{\mu\nu}$ in the Landau frame
\begin{equation}
 T^{\mu\nu} = \ped \, U^{\mu} U^{\nu} -\left(\vp\pres + \Pi \right) \Delta^{\mu\nu} +\pi^{\mu\nu},
\end{equation}
where $\Pi$ is the bulk pressure, $\pi^{\mu\nu}$ is the shear stress tensor (the space-like, symmetric, and traceless part of $T^{\mu\nu}$), and the projector $\Delta^{\mu\nu}$ reads
\begin{equation}
 \Delta^{\mu\nu} = g^{\mu\nu} - U^\mu U^\nu.
 \label{Delta}
\end{equation}
The distinction between the hydrostatic (equilibrium) pressure $\pres$ and the bulk pressure $\Pi$ requires a thermodynamic input. A very common procedure is to define an effective (point-dependent) temperature through the Landau matching
\begin{equation}
 \ped(x) = \pedeq\left(\vp T(x)\right),
 \label{Landau_matching}
\end{equation}
where $\pedeq(T)$ is the energy density of the system at global equilibrium in the thermodynamic limit. In this way, the hydrostatic pressure is defined through the effective temperature, as the pressure that the system would have in thermodynamical equilibrium with temperature~$T$, namely
\begin{equation}
 \pres(x) = \preseq\left(\vp T(x)\right).
\end{equation}
Since we consider, for mathematical simplicity, only the case of a conformal fluid, the equation of state takes the form
\begin{equation}
 \pres(x) = \frac{1}{3}\ped(x),
 \label{EOS}
\end{equation}
and the bulk viscosity $\Pi$ vanishes
\begin{equation}\label{conf}
 \Pi = 0.
\end{equation}
Contracting~(\ref{en_mom})  with the four-velocity $U_\nu$ and the projector $\Delta^\alpha_\nu$, respectively, and taking into account the conformal requirement~(\ref{conf}),  after some straightforward algebra one obtains the equations
\begin{equation}
 D \ped = -\left(\vp\ped + \pres  \right) \theta +\pi_{\mu\nu} \sigma^{\mu\nu},
 \label{en_con_0}
\end{equation}
and
\begin{equation}
 \left( \enlpar \ped + \pres \right) DU^\alpha = \nabla^\alpha\pres  - \Delta^\alpha_\mu \partial_\nu \pi^{\mu\nu},
 \label{mom_con_0}
\end{equation}
with $D = U^\mu\partial_\mu$ being the convective derivative, $\theta=\partial_\mu U^\mu$ the  expansion scalar, $\nabla^\mu=\Delta^{\mu\nu}\partial_\nu$ the spatial gradient, and $\sigma^{\mu\nu}$ the shear flow tensor
\begin{equation}
 \sigma^{\mu\nu} = \partial^{\langle \mu} U^{\nu\rangle}.
\end{equation}
Here we make use of the notation
\begin{eqnarray}
 {\sf A}^{\langle \mu\nu \rangle} &=& 
 \Delta^{\mu\nu}_{\alpha\beta} \; {\sf A}^{\alpha \beta}, \nonumber
\end{eqnarray}
where
\begin{eqnarray} 
 \Delta^{\mu\nu}_{\alpha\beta}  &=&  \frac{1}{2}\left( \vphantom{\frac{}{}} \Delta^\mu_\alpha \Delta^\nu_\beta + \Delta^\nu_\alpha \Delta^\mu_\beta -\frac{2}{3}\Delta^{\mu\nu} \Delta_{\alpha\beta} \right).\nonumber
\end{eqnarray}
%
\section{Implementation of (1+1)D non-boost-invariant expansion}
\label{sect:implement}
%
Due to translational invariance in the transverse plane of the systems we study in this work, the considered evolution of matter becomes effectively (1+1)-dimensional [(1+1)D]. Hence, each scalar quantity must depend only on the lab frame time $t$ and on the lab frame longitudinal direction $z$. It is convenient, however,  to change at this point to the (longitudinal) proper time $\tau$ and the spacetime rapidity~$\eta$,
\begin{equation}
 \tau = \sqrt{ t^2 - z^2 }, \qquad \qquad \eta = \frac{1}{2} \ln \left( \frac{t+z}{t-z} \right),
\end{equation}
where $t>|z|$. Because of the rotational and translational invariance in the transverse plane, the four-velocity vector $U^\mu$ must have vanishing components in the transverse direction. Therefore, the four-velocity of the fluid can be written as
\begin{equation}
 U^\mu = \left(\vp\cosh(\eta + \thetapar), 0 , 0, \sinh(\eta + \thetapar) \right),
 \label{U}
\end{equation}
with $\thetapar(\tau,\eta)$ being a scalar function. In the limit $\thetapar = 0$, we recover the boost-invariant four-velocity profile. It is convenient now to fix a complete orthonormal basis~\cite{Florkowski:2011jg,Martinez:2012tu,Tinti:2013vba}. In addition to the time-like four-velocity vector $U^\mu$, we introduce the longitudinal-direction four-vector
\begin{equation}
 Z^\mu = \left(\vp\sinh(\eta + \thetapar), 0 , 0, \cosh(\eta + \thetapar) \right),
 \label{Z}
\end{equation}
and two arbitrary four-vectors in the transverse plane, $X^\mu$ and $Y^\mu$, which can be, without loss of generality, identified with the unit vectors along the $x$ and $y$ directions in the lab frame. In the local rest frame (LRF) the basis vectors read
\begin{eqnarray}
&& U^\mu = (1,0,0,0),\quad X^\mu = (0,1,0,0),\nonumber \\
&& Y^\mu = (0,0,1,0),\quad Z^\mu = (0,0,0,1).
\end{eqnarray}
We can therefore write the projector~(\ref{Delta}) in the following way
\begin{eqnarray}
 \Delta^{\mu\nu} &=& g^{\mu\nu} - U^\mu U^\nu  =  - X^\mu X^\nu -Y^\mu Y^\nu- Z^\mu Z^\nu 
 \nonumber \\
 &=& -\sum_I I^\mu I^\nu,
\end{eqnarray}
where the sum, hereafter, is meant to run over the space-like basis vectors, $I = X, Y, Z$. 

The directional derivatives read
\begin{eqnarray}\nonumber
&& D = U^\mu \partial_\mu = \cosh\thetapar \partial_\tau +\frac{\sinh\thetapar}{\tau}\partial_\eta, \\ \nonumber  
&& X^\mu\partial_\mu = \partial_x, \quad Y^\mu\partial_\mu = \partial_y, \\
&& D_L = Z^\mu\partial_\mu = \sinh\thetapar \partial_\tau +\frac{\cosh\thetapar}{\tau}\partial_\eta.
\label{dir_der}
\end{eqnarray}
Making use of~(\ref{U}) and~(\ref{dir_der}), the four-acceleration $D U^\mu$ reads
\begin{equation}
 DU^\mu = \left( D\thetapar  + \frac{\sinh\thetapar}{\tau}\right)Z^\mu.
 \label{acc}
\end{equation}
We note that in the limit $\thetapar \to 0$ we recover the  Bjorken result, $DU^\mu=0$. In the next step, we construct
the expansion tensor 
\begin{equation}
 \theta^{\mu\nu} = \frac{1}{2}\Delta^{\mu\alpha}\Delta^{ \nu\beta} \left( \enlpar \partial_\alpha U_\beta + \partial_\beta U_\alpha \right) 
 = -\sum_{I}\theta_I \,  I^\mu I^\nu,
\end{equation}
where
\begin{equation}
 \theta_I = U_\mu \vp (I^\nu \partial_\nu) I^\mu  =
 \begin{cases}
   \frac{ \cosh\thetapar}{\tau}+ D_L \thetapar & I=Z, \\
   0 & I=X,Y.
 \end{cases}
\end{equation}
Then, we find the expansion scalar 
\begin{eqnarray}
 \theta &=& \partial_\mu U^\mu = \Delta^{\mu\nu} \partial_\mu U_\nu  \nonumber \\
 &=& \Delta^{\mu\nu} \theta_{\mu\nu} = \theta_Z = \frac{\cosh\thetapar}{\tau}+ D_L \thetapar,
\end{eqnarray}
and the shear flow tensor 
\begin{equation}
 \sigma^{\mu\nu} = \theta^{\mu\nu} -\frac{\theta}{3}  \Delta^{\mu\nu} = \sum_I \sigma_I \, I^\mu I^\nu,
\end{equation}
with
\begin{equation}
\sigma_I =  \frac{\theta}{3} -\theta_I =
\begin{cases}
 -\frac{2}{3}\theta & I=Z, \\
 \hphantom{-}\frac{1}{3}\theta & I=X,Y.
\end{cases} 
\end{equation}
We note that the vorticity $\omega^{\mu\nu}$ vanishes for the type of expansion considered in this work
\begin{equation}
\omega^{\mu\nu} = \frac{1}{2}\Delta^{\mu\alpha}\Delta^{\nu\beta}\left( \enlpar \partial_\alpha U_\beta -\partial_\beta U_\alpha \right) = 0.
\end{equation}

The forms of the four-acceleration~(\ref{acc}) and the directional derivatives~(\ref{dir_der}), as well as the symmetry in the transverse plane, imply that only the $Z_\alpha$ projection of Eq.~(\ref{mom_con_0}) is not trivially satisfied, namely
\begin{equation}
 -\left( \enlpar \ped + \pres \right)\left( D\thetapar  + \frac{\sinh\thetapar}{\tau}\right) = D_L  \pres  - Z_\mu \partial_\nu \pi^{\mu\nu}.
 \label{mom_cons_1}
\end{equation}
The symmetry of the system constrains further possible forms of $\pi^{\mu\nu}$. Indeed, in the $\{U, X, Y, Z\}$ basis, the tensor $\pi^{\mu\nu}$ must be diagonal and with equal values of the $X$ and $Y$ components
\begin{eqnarray}
 \pi^{\mu\nu} &=& \sum_I \pi_I  \, I^\mu I^\nu \\
& = & \frac{\pi_s(\tau,\eta)}{2}\,   \, \left( \enlpar  X^\mu X^\nu + Y^\mu Y^\nu \right) -\pi_s(\tau,\eta) \, Z^\mu Z^\nu.
\nonumber
\end{eqnarray}
Here $\pi_X = \pi_Y =  \pi_s/2$ and $\pi_Z = -\pi_s$. The scalar $\pi_s$ completely defines the shear pressure corrections in the (1+1)D expansion.

Besides the components of the shear stress tensor, it is convenient to introduce the transverse and longitudinal pressure\footnote{We make use of the conformal equation of state. In the non-conformal case one has to include the bulk pressure $\Pi$ in the two definitions}, namely $\pres_T$ and $\pres_L$, that are defined by the expressions
\begin{equation}
 \pres_T =X_\mu X_\nu T^{\mu\nu} = Y_\mu Y_\nu T^{\mu\nu} = \pres +\frac{\pi_s}{2},
 \label{PT}
\end{equation}
\begin{equation}
 \pres_L = Z_\mu Z_\nu T^{\mu\nu} = \pres -\pi_s.
 \label{P_L}
\end{equation}
In this way, Eqs.~(\ref{en_con_0}) and~(\ref{mom_cons_1}) reduce to the compact form
\begin{equation}
 D\ped = -\left( \enlpar \ped + \pres_L \right) \theta, \qquad \;
D_L \pres_L = -\left( \enlpar \ped + \pres_L \right) \theta_L,
 \label{en_mom_con_0}
\end{equation}
where the divergence of the vector $Z^\mu$ reads
\begin{eqnarray}
\theta_L &=& \partial_\mu Z^\mu = g^{\mu\nu}\partial_\mu Z_\nu = \left(\enlpar U^\mu U^\nu + \Delta^{\mu\nu} \right) \partial_\mu Z_\nu
 \nonumber \\
&=& U^\mu D Z_\mu = -Z_\mu DU^\mu = D\thetapar  + \frac{\sinh\thetapar}{\tau}.
\end{eqnarray}

Since in viscous hydrodynamics the shear pressure $\pi^{\mu\nu}$ (and bulk pressure $\Pi$ in the non-conformal case) corrections are treated as independent variables, it is convenient to explicitly maintain these variables in Eqs.~(\ref{en_mom_con_0}),
\begin{eqnarray} 
 D\ped &=& -\left( \enlpar \ped + \pres  - \pi_s \right) \theta,   \label{en_mom_con_IS1}\\
D_L \left( \enlpar \pres  -\pi_s \right) &=& -\left( \enlpar \ped +  \pres  -\pi_s \right) \theta_L.
 \label{en_mom_con_IS2}
\end{eqnarray}
%

\section{Second-order viscous hydrodynamics}
\label{sect:secondorder}

The system of equations (\ref{en_mom_con_IS1}) and (\ref{en_mom_con_IS2}) is clearly not closed, having two independent equations for three independent variables:  the effective temperature $T$ (giving both the energy density $\ped$ and the hydrostatic pressure $\pres$ through Eqs.~(\ref{Landau_matching}) and~(\ref{EOS})), the  relative fluid rapidity $\thetapar$ (the only independent component of the four-velocity), and $\pi_s$ (expressing the shear pressure correction). In second-order viscous hydrodynamics, this system is closed with additional dynamic equations for the shear stress tensor which is treated as a new hydrodynamic variable.

\subsection{M\"uller-Israel-Stewart (MIS) approach}

Since the original works by M\"uller in 1967~\cite{Muller:1967zza} and Israel and Stewart in 1976~\cite{Israel:1976tn,Israel:1979wp}, second-order hydrodynamics has been gradually evolving into a more and more complete theory that includes various terms determining the dynamics of the bulk pressure and the shear stress tensor. One popular version of this theory, used in many phenomenological applications and denoted below as MIS, is based on the formula \cite{Muronga:2001zk,Heinz:2005bw,Bozek:2009dw}
\begin{eqnarray}
 D\pi^{\langle\mu\nu\rangle} +\frac{\pi^{\mu\nu}}{\tau_\pi}  &=& 2\beta_\pi \sigma^{\mu\nu}  - \pi^{\mu\nu} \frac{T \beta_\pi}{2}\partial_\rho \left( \frac{1}{T\beta_\pi} U^\rho \right). \nonumber \\ \label{shear}
\end{eqnarray}
Here  $\tau_\pi$ is the shear relaxation time, while $\beta_\pi = \eta/\tau_\pi$ is the ratio of the first-order transport coefficient and its respective relaxation time.~\footnote{Note that we use the same symbol for the shear viscosity and the rapidity, however, its meaning should be always clear from the context.} In the particular case of the (1+1)D non-boost-invariant expansion considered in this work, Eqs.~(\ref{shear}) reduce, as expected, to three degenerate equations for $\pi_I$'s
\begin{equation}
 D\pi_I +\frac{\pi_I}{\tau_\pi}  =  2\beta_\pi \sigma_I  - \pi_I \frac{T \beta_\pi}{2}\partial_\mu \left( \frac{1}{T\beta_\pi} U^\mu \right),
 \label{pi_I}
\end{equation}
or, equivalently, to a single equation for $\pi_s$
\begin{equation}
 D\pi_s +\frac{\pi_s}{\tau_\pi}  = \frac{4}{3}\beta_\pi \theta  - \pi_s \frac{T \beta_\pi}{2}\partial_\mu \left( \frac{1}{T\beta_\pi} U^\mu \right).
 \label{pi_s}
\end{equation}
A closer examination of Eq.~(\ref{pi_s}) shows  that it includes higher-order terms in gradients (see App.~\ref{sect:thirdorder} for the details). If they are neglected, Eq.~(\ref{pi_s}) reduces to the form
\begin{eqnarray}
D\pi_s+\frac{\pi_s}{\tau_\pi}= \frac{4 \eta}{3\tau_\pi}   \theta - \frac{4}{3} \theta \pi_s.   
 \label{pi_s_MIS}
\end{eqnarray}
Summing up, given the symmetry constraints, the three MIS equations for the hydrodynamical evolution are the energy and momentum conservation equations~(\ref{en_mom_con_IS1})-(\ref{en_mom_con_IS2}), and the relaxation-type equation (\ref{pi_s_MIS}).

\subsection{Denicol-Niemi-Molnar-Rischke (DNMR) approach}
In the recent works~\cite{Denicol:2010xn,Denicol:2012cn,Denicol:2014loa}, starting from kinetic theory, Denicol, Niemi, Molnar and Rischke have derived equations for viscous hydrodynamics in an expansion controlled by the Knudsen number and the inverse Reynolds number. In the conformal case and the relaxation time approximation (RTA) for the collisional kernel, this approach leads to the evolution equations which are equivalent to the ones resulting from the modified Chapman-Enskog prescription proposed by Jaiswal \cite{Jaiswal:2013npa,Jaiswal:2013vta}.  The situation in the DNMR case is similar to that seen previously for MIS, however, some terms appearing in the final DNMR equation for $\pi_s$ are different from those used in MIS, namely, the DNMR approach in the conformal limit, with vorticity neglected once again, yields
\begin{equation}
 D\pi_s + \frac{\pi_s}{ \tau_\pi}  = \frac{4\eta}{3 \tau_\pi} \, \theta  - \delta_{\pi\pi} \, \theta \, \pi_s - \frac{1}{3}\theta\, \tau_{\pi\pi}\, \pi_s 
 +\frac{\phi_7}{2 \tau_\pi} \, \pi_s^2.
 \label{pi_s_DNMR}
\end{equation}
The quantities $\delta_{\pi\pi}$, $\tau_{\pi\pi}$, and $\phi_7$ are second-order coefficients. In order to compare uniformly various approaches we use the values obtained from the RTA kinetic theory in the conformal case:
\begin{eqnarray}
\delta_{\pi\pi} = \frac{4}{3}, \quad \tau_{\pi\pi} = \frac{10}{7}, \quad \phi_7 = 0.
\end{eqnarray}
This allows us to simplify (\ref{pi_s_DNMR}) to
\begin{equation}
 D\pi_s + \frac{\pi_s}{ \tau_\pi}  = \frac{4\eta}{3 \tau_\pi} \, \theta  - \frac{38}{21} \, \theta \, \pi_s.
  \label{pi_s_DNMR_1}
\end{equation}

\subsection{Baier-Romatschke-Son-Starinets-Stephanov (BRSSS) approach}

The BRSSS approach \cite{Baier:2007ix} uses arguments of conformal symmetry to construct the shear stress tensor out of gradients of the hydrodynamic variables $T$ and $U^\mu$ (up to the second order in gradients). To obtain the dynamic equation for $\pi^{\mu\nu}$ the Navier-Stokes (NS) equations are used. This procedure leads to a structure similar to (\ref{pi_s_DNMR}), namely
\begin{equation}
 D\pi_s + \frac{\pi_s}{ \tau_\pi}  = \frac{4 \eta}{3  \tau_\pi} \, \theta  - \frac{4}{3} \, \theta \, \pi_s  -\frac{ \lambda_1}{2  \tau_\pi \eta^2} \, \pi_s^2,
 \label{pi_s_BRSSS}
\end{equation}
where \cite{Romatschke:2011qp}
\begin{eqnarray}
\lambda_1 =\frac{5 \eta \tau_\pi}{7}.
\end{eqnarray}

\section{Anisotropic hydrodynamics formalism}
\label{sect:ahydro}
%
Equations~(\ref{en_mom_con_0}) must be fulfilled by any hydrodynamic approach, since they describe energy and momentum conservation. Within the anisotropic hydrodynamics framework, we express the energy density, the longitudinal pressure, and the transverse pressure by the formulas
\begin{eqnarray}
 \ped &=& \pedeq(\Lambda){\cal R} (\xi)= 3 \preseq (\Lambda) {\cal R}({\xi}), \nonumber \\
 \pres_L &=& \preseq(\Lambda) {\cal R}_L(\xi), \nonumber \\
 \pres_T &=& \preseq(\Lambda) {\cal R}_T(\xi),
\end{eqnarray}
where the functions $\cal R$, ${\cal R}_L$, and ${\cal R}_T$ have been defined in Ref.~\cite{Martinez:2010sc}. The quantity $\Lambda$ is the transverse momentum scale, while $\xi$ is the anisotropy parameter. In the limit $\xi \to 0$, the transverse momentum scale $\Lambda$ may be identified with the temperature of the system in local equilibrium.

The two non-trivial equations~(\ref{en_mom_con_0}) that follow from the energy-momentum conservation can be written in this case as
\begin{eqnarray}
&& {\cal R}(\xi) D\ln\preseq(\Lambda) + {\cal R} ^\prime(\xi)  D\xi  \nonumber \\
&& \hspace{2cm} +\frac{2}{3}\left[ \vphantom{\frac{}{}} {\cal R}_T(\xi)  +{\cal R}_L(\xi)\right]\theta = 0 
\label{emca1c} \\
&&{\cal R }_L(\xi) D_L\ln\preseq(\Lambda) + {\cal R}_L^\prime(\xi)  D_L\xi \nonumber \\
&& \hspace{2cm} +2\left[\vphantom{\frac{}{}} {\cal R}_T(\xi)  + {\cal R}_L(\xi)\right]\theta_L = 0.
\label{emca2c}   
\end{eqnarray}
The primes in  ${\cal R}^\prime$ and ${\cal R}^\prime_L$ denote the partial derivative with respect to the argument $\xi$.

It has been demonstrated in Ref.~\cite{Tinti:2015xwa} that extra dynamical equations for anisotropic hydrodynamics can be most optimally derived from the exact equation for the shear stress tensor, which is obtained directly from the Boltzmann equation (this procedure requires however a truncation of the expansion of the phase-space distribution function and keeping the leading anisotropic term only). The method proposed in Ref.~\cite{Tinti:2015xwa}  yields the best agreement with the exact solutions of the Boltzmann equation in the case of a one-dimensional, longitudinally boost-invariant flow (although the very same method can be applied to derive anisotropic hydrodynamics equations in a general (3+1)D case). 

In this work we use the results obtained in \cite{Tinti:2015xwa}. It is straightforward to check that for anisotropic hydrodynamics the five independent equations for the shear stress tensor reduce to a single independent equation for $\pi_s = 2(\pres_T-\pres_L)/3$, in the similar way as it happens in viscous hydrodynamics. Thus, one can take, without loss of generality, the longitudinal projection of Eq.~(40) in Ref.~\cite{Tinti:2015xwa} as a supplementary equation. The latter reads
\begin{eqnarray}
&&  D(\pres_T -\pres_L) +\frac{1}{\taueq}(\pres_T-\pres_L) \nonumber \\
&& =  \theta \preseq (\Lambda)\left\{ 2\sqrt{1+\xi} \, \partial_\xi \left[ \sqrt{1+\xi}\, {\cal R}_T(\xi)\right]  \right. \nonumber \\
&& \left. - \frac{2}{\sqrt{1+\xi}}\partial_\xi \left[ \left( \sqrt{1+\xi} \right)^3 {\cal R}_L(\xi)\right] \right\} \\ \nonumber
&& \qquad \qquad  - \theta(\pres_T -\pres_L) + 2 \theta\pres_L,
\end{eqnarray}
which can be further simplified dividing both sides by the positive quantity $\preseq$. After some algebra, the last equation can be rewritten as
\begin{eqnarray}
&&\left[ \vphantom{\frac{}{}} {\cal R}_T(\xi)  -{\cal R}_L(\xi)\right ]\left[ D\ln\preseq(\Lambda) +\frac{1}{\tau_{\rm eq}}\right] \nonumber \\
&&  + \left[ \vphantom{\frac{}{}}{\cal R}_T ^\prime(\xi) - {\cal R}_L^\prime(\xi)\right]  \left[ D\xi   - 2 (1+\xi)  \theta\vp\right] = 0. \label{xiac}  
\end{eqnarray}

\begin{figure*}[h!]
\begin{center}
\includegraphics[width=0.9 \textwidth]{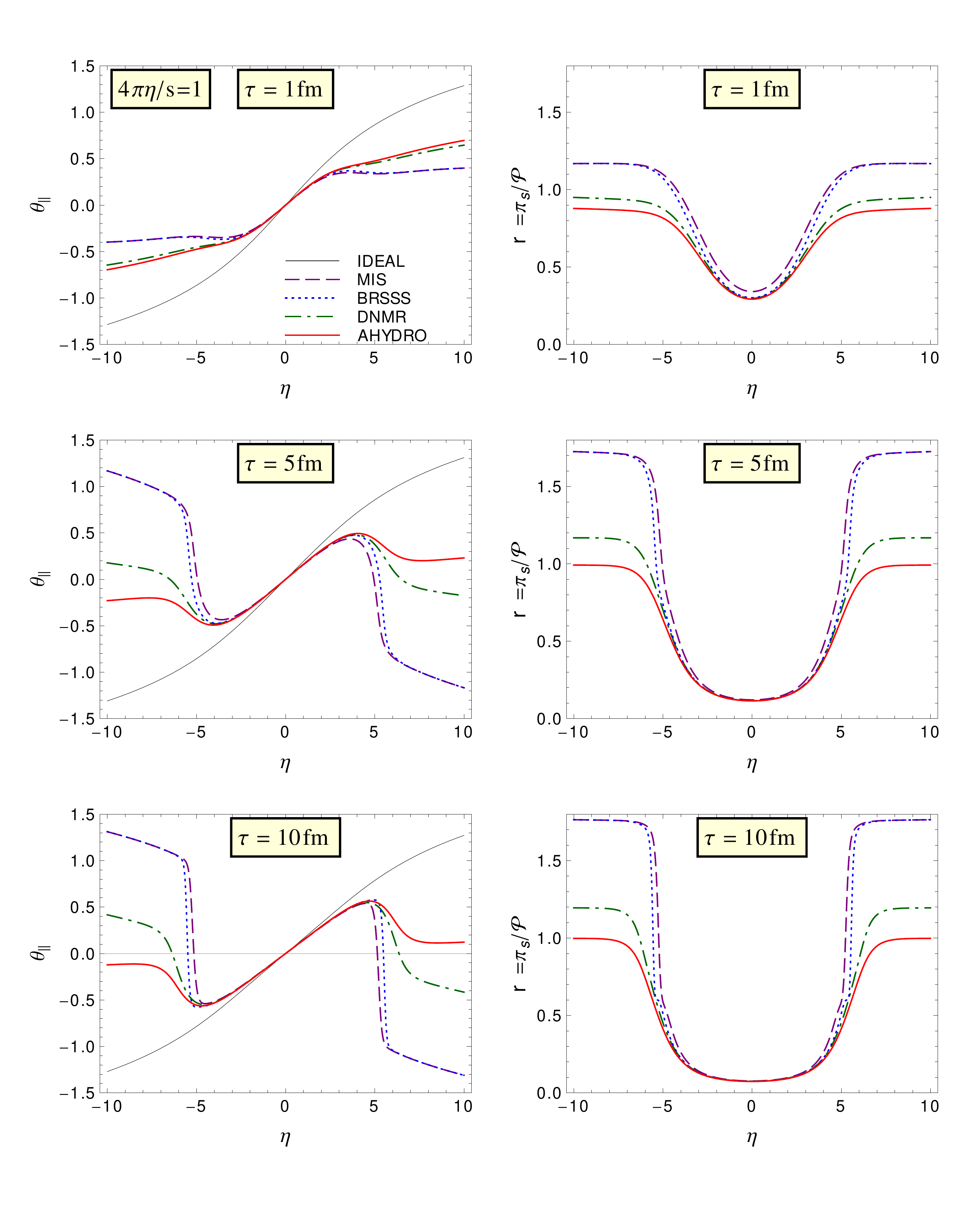}
\end{center}
\caption{(Color online) Pseudorapidity profiles of the relative flow rapidity $\theta_\parallel$ (left) and the ratio $r=\pi_s/{\cal P}$  (right)  at the proper time $\tau=1$ fm (top), $\tau=5$ fm (middle), and $\tau=10$ fm (bottom) obtained within perfect fluid hydrodynamics (thin solid black lines), MIS (dashed purple lines),  BRSSS (dotted blue lines), DNMR (dashed dotted green lines) and AHYDRO (solid red lines) with $\bar{\eta}=1/(4\pi)$.}
\label{fig:1over4pi}
\end{figure*}

\begin{figure*}[t!]
\begin{center}
\includegraphics[width=0.9 \textwidth]{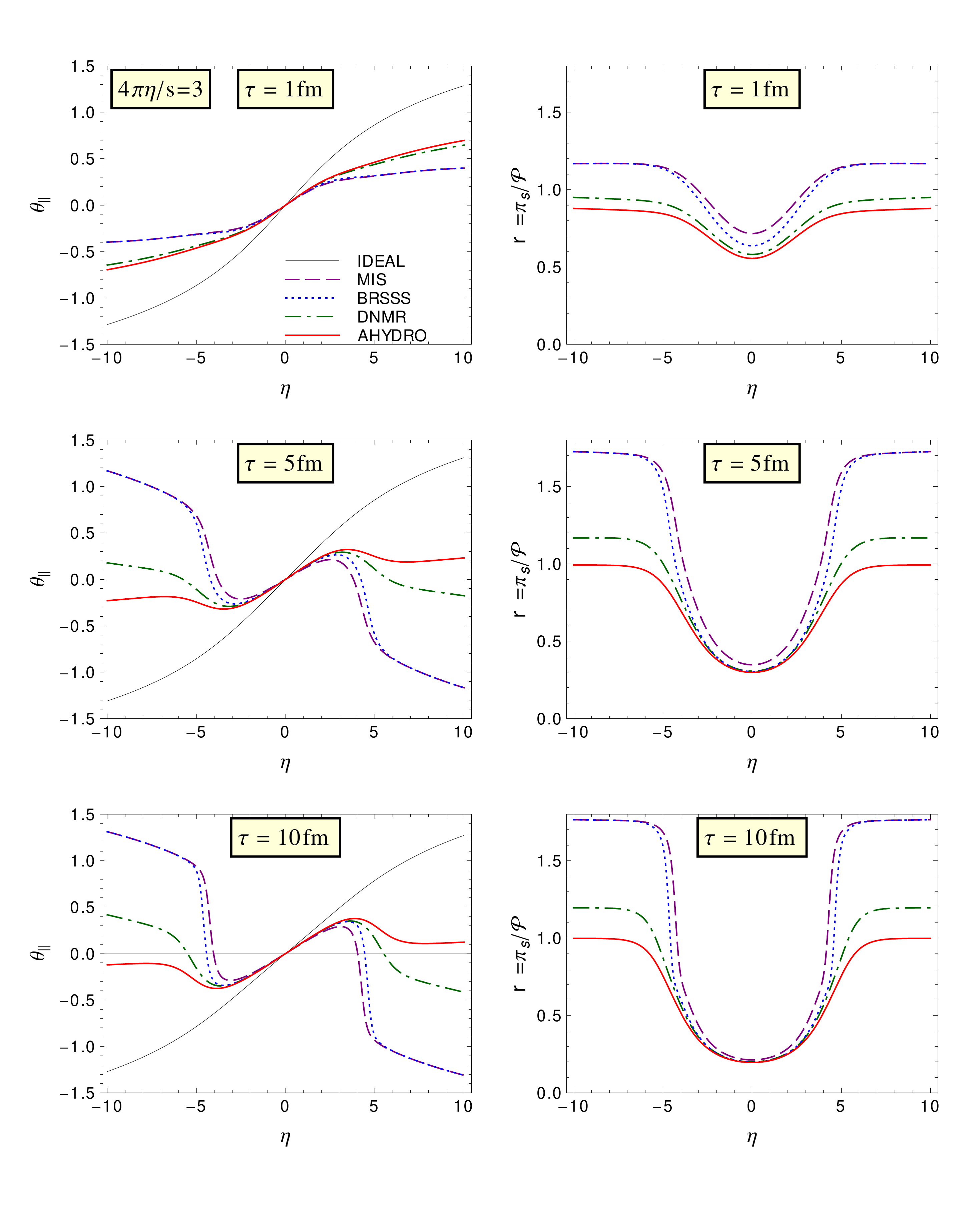}
\end{center}
\caption{(Color online) Same as Fig.~\ref{fig:1over4pi} but with $\bar{\eta}=3/(4\pi)$.}
\label{fig:3over4pi}
\end{figure*}

\begin{figure}[t]
\begin{center}
\includegraphics[width=0.98 \columnwidth]{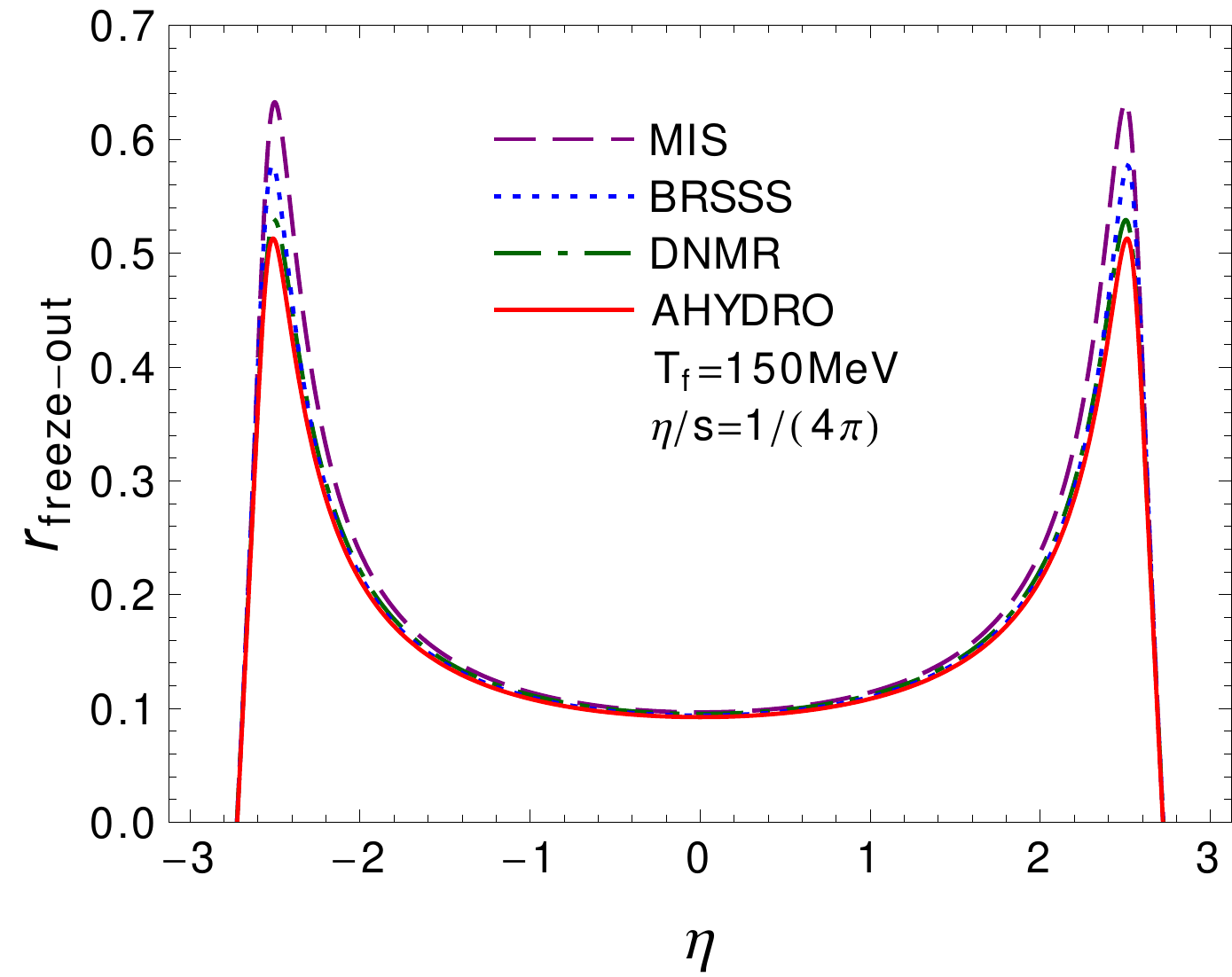}\\
\end{center}
\caption{(Color online) Ratio $\pi_s/{\cal P}$ calculated for the case $4\pi{\bar \eta} = 1$ along the freeze-out hypersurface of constant temperature of $T=0.15$ GeV. }
\label{fig:1over4piFO}
\end{figure}

\begin{figure}[t]
\begin{center}
\includegraphics[width=0.98 \columnwidth]{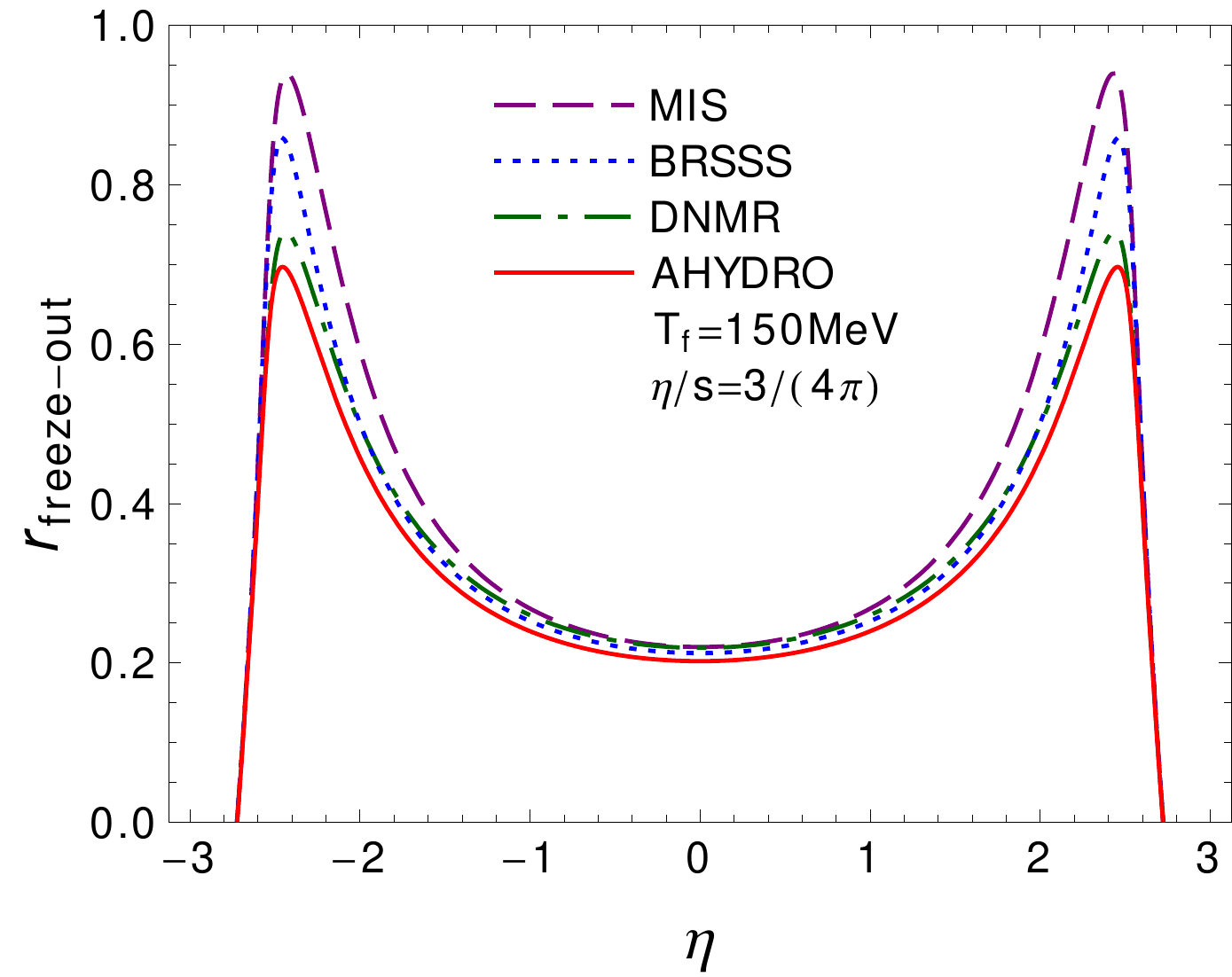}\\
\end{center}
\caption{(Color online) Same as Fig.~\ref{fig:1over4piFO} but for $4\pi{\bar \eta} = 3$.}
\label{fig:3over4piFO}
\end{figure}

\begin{figure}[t]
\begin{center}
\includegraphics[width=0.98 \columnwidth]{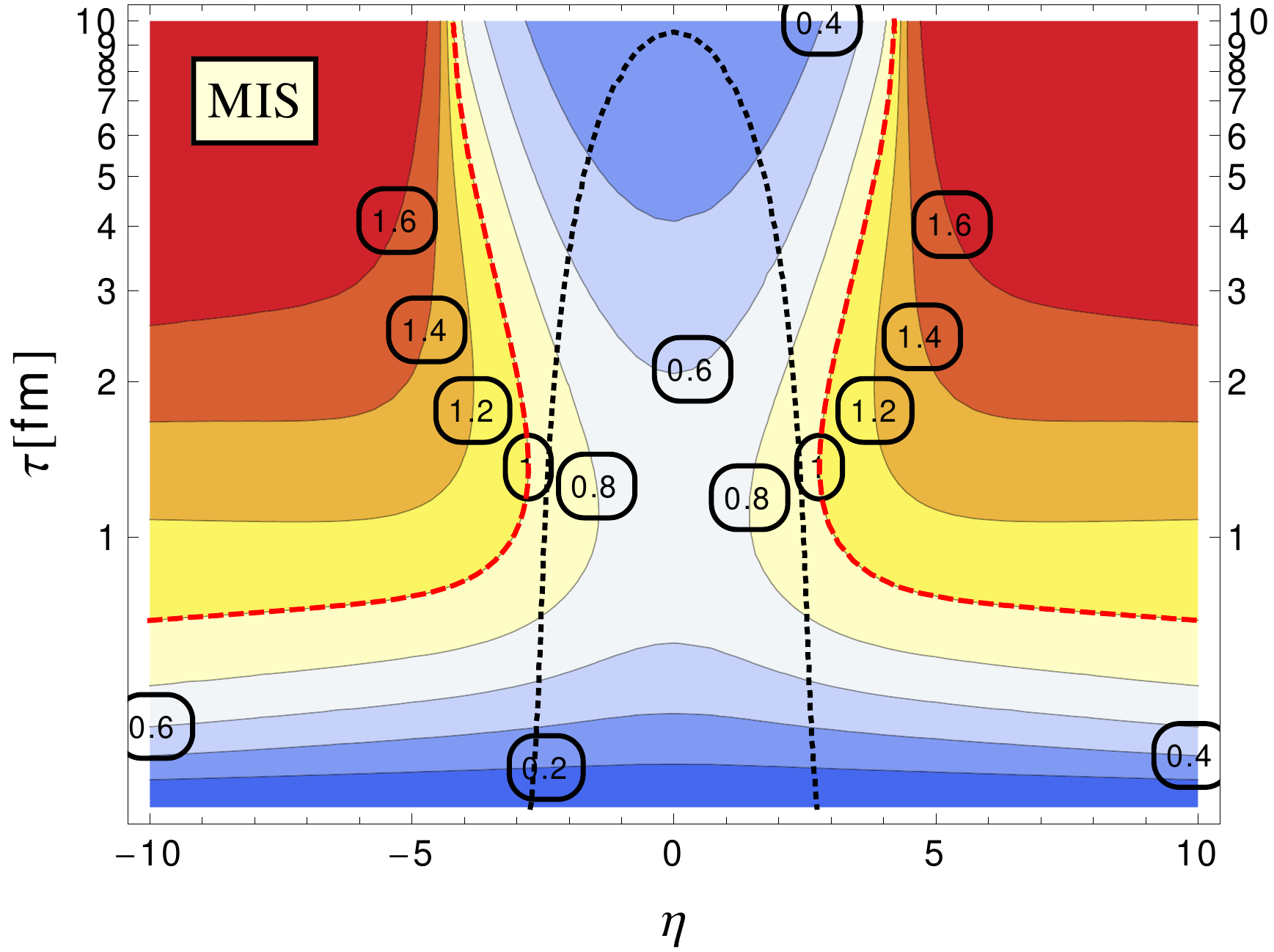}\\
\end{center}
\caption{(Color online) Freeze-out curve of $T=0.15$ GeV (dotted black line) shown in the space-time diagram of $r(\tau,\eta)$ for MIS with $4\pi{\bar \eta} = 3$.
}
\label{fig:r_MIS_etabar3}
\end{figure}

\begin{figure}[t]
\begin{center}
\includegraphics[width=0.98 \columnwidth]{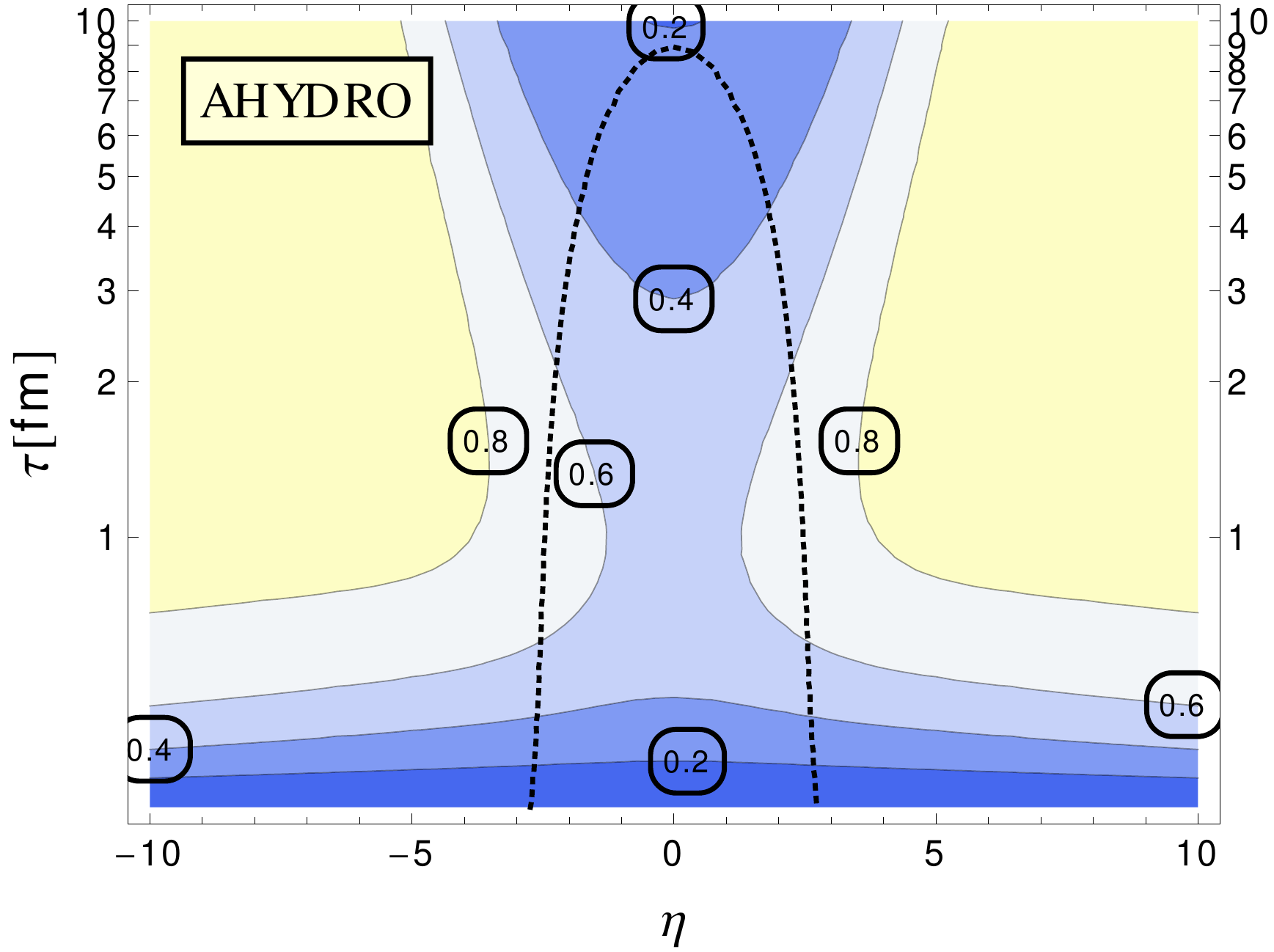}\\
\end{center}
\caption{ (Color online) Same as Fig.~\ref{fig:r_MIS_etabar3} but for AHYDRO.
}
\label{fig:r_AH_etabar3}
\end{figure}

\section{Numerical results}
\label{sect:results}

Dealing with the conformal case, one obtains three independent equations, for example,  for $T$, $\thetapar$, and $\pi_s$ (in the case of viscous hydrodynamics) or for $\Lambda$, $\thetapar$, and $\xi$ (for anisotropic hydrodynamics). The appropriate selection of equations is: (\ref{en_mom_con_IS1}), (\ref{en_mom_con_IS2}) and (\ref{pi_s_MIS}) for MIS; (\ref{en_mom_con_IS1}), (\ref{en_mom_con_IS2}) and (\ref{pi_s_DNMR_1}) for DNMR; (\ref{en_mom_con_IS1}), (\ref{en_mom_con_IS2}) and (\ref{pi_s_BRSSS}) for BRSSS, and, finally, (\ref{emca1c}),  (\ref{emca2c}), and (\ref{xiac}) for AHYDRO.
 
In all cases the relaxation time depends inversely on the temperature  $\tau_{\pi}=\tau_{\rm eq} = 5 {\bar \eta}/T$, where ${\bar \eta} = \eta/s$ is the shear viscosity to entropy density ratio. We use the values $4\pi{\bar \eta} = 1$ and  $4\pi{\bar \eta} = 3$, which correspond to the range of present experimental estimates. We initialize the energy density with a Gaussian space-time rapidity profile ${\cal E}(\tau_0,\eta)={\cal E}_0 \exp\left(-  \eta^2 /(2 a^2)\right)$, where $a =0.9$ and ${\cal E}_0=100$~GeV$/{\rm fm}^3$ at the initial proper time $\tau_0=0.3$~fm.  The initial pressure is taken in the equilibrium (conformal) form $\pres(\tau_0,\eta)= {\cal E}(\tau_0,\eta)/3$. The temperature is obtained from the relation $\alpha\, T^4 = {\cal E}$, where $\alpha = 15.63$. The initial longitudinal rapidity is  very close to the Bjorken formula, namely,  $\thetapar (\tau_0,\eta)=  b \eta$ with $b=10^{-10}$. We start with a practically isotropic pressure in the system, taking $\pi_s(\tau_0,\eta) = c \pres(\tau_0,\eta)$ where $c=10^{-10}$. The initial anisotropy parameter for anisotropic hydrodynamics calculations is extracted from the equation $\pi_s(\tau_0,\eta)/\pres(\tau_0,\eta) = 1-{\cal R}_L(\xi(\tau_0,\eta))/{\cal R}(\xi(\tau_0,\eta))$.

In Figs.~\ref{fig:1over4pi} and ~\ref{fig:3over4pi} we present the relative fluid rapidity $\theta_\parallel$ (left) and the ratio $r=\pi_s/{\cal P}$ (right) as functions of spacetime rapidity at different values of the proper time, $\tau=1$~fm (top), $\tau=5$~fm (middle), and $\tau=10$~fm (bottom), obtained within perfect-fluid hydrodynamics (thin solid black lines), MIS (dashed purple lines),  BRSSS (dotted blue lines), DNMR (dashed dotted green lines), and AHYDRO (solid red lines).  In Fig.~\ref{fig:1over4pi} we use the lower bound value of $\bar{\eta}=1/(4\pi)$, while in Fig.~\ref{fig:3over4pi} the value $\bar{\eta}=3/(4\pi)$ is used.  

In the central rapidity region we observe that matter is accelerated outwards in the forward/backward rapidity directions during the entire considered evolution time. This property is qualitatively similar for all considered formalisms. On the other hand, at large values of rapidity, $|\eta| > 5$, due to long relaxation times (being inversely proportional to the temperature of the system, $\tau_{\rm eq} \sim 1/T$), large dissipative corrections are quite quickly being built up. In the various dissipative hydrodynamics formalisms these large dissipative corrections are handled in different ways, according to the specific evolution equations used for $\pi_s$. It turns out that in all approaches, except for AHYDRO, the dissipative corrections lead to the values of $r=\pi_s/{\cal P}$ exceeding unity. One should stress here that, according to Eq.~(\ref{P_L}), in second-order viscous hydrodynamics having $r>1$ is equivalent to the generation of a negative longitudinal pressure in the system. This situation has no physical interpretation in the kinetic theory background, which all formalisms considered are based on, unless there are fields in the system (which are explicitly neglected in all cases). By construction, the AHYDRO prescription does not allow for negative pressures, thus it regulates dissipative phenomena at large values of $\eta$.

In all formalisms considered, the phenomena taking place at small and large spacetime rapidities lead eventually to the creation of a shock-wave in both $\theta_{\parallel}$ and $\pi_s$. This shockwave can cause problems in numerical simulations due to the large spatial gradients induced.  As a result, for our numerical simulations we use the weighted LAX (wLAX) algorithm with $\lambda_{\rm LAX} = 0.01$ to handle spurious oscillations arising when using center-differences schemes in high-gradient regions~\cite{Martinez:2012tu}.\footnote{We note here that, without wLAX smoothing, all codes, except for AHYDRO, crash within the evolution time.}  For temporal updates, we use standard fourth-order Runge-Kutta.  We would like to stress here that although the shock-waves are sometimes expected natural phenomena, their creation in the presented results strongly depends on the formalism used. Therefore, their creation may affect the interpretation of the final results and increase numerical difficulties, especially if event-by-event fluctuations in rapidity are considered. Based on the observed behavior, we may state that the time needed for creation of a shock-wave  is ordered as follows: $\tau_{\rm shock}^{\rm BRSSS}< \tau_{\rm shock}^{\rm MIS}< \tau_{\rm shock}^{\rm DNMR}~\ll~\tau_{\rm shock}^{\rm AHYDRO}$. Moreover, comparing Figs.~\ref{fig:1over4pi} and ~\ref{fig:3over4pi}  we find that larger times are required to build up a shock-wave in the case with larger viscosity.  

In Figs.~\ref{fig:1over4piFO} and \ref{fig:3over4piFO} we show the ratio $\pi_s/{\cal P}$ calculated along the freeze-out curve of constant temperature of $T=0.15$ GeV (by the freeze-out curve we mean the projection of the freeze-out 3D hypersurface on the plane with fixed transverse coordinates). We observe differences between the values of $r$ obtained with different hydrodynamics frameworks at large rapidities. Such differences grow with increasing viscosity. We note that differences in the freeze-out curves introduce uncertainties in the spectra of the particles emitted at large rapidities.

It is interesting to plot the freeze-out curves in the space-time diagram spanned by the $\eta$ and $\tau$ coordinates, which is shown for the case $4\pi{\bar \eta} = 3$ in Figs.~\ref{fig:r_MIS_etabar3} and \ref{fig:r_AH_etabar3} for MIS and AHYDRO, respectively. One can notice that the extracted freeze-out curve for MIS goes very closely to the region with unphysical negative pressure; note the red dashed lines corresponding to the boundary value where $r=1$. In this case, small perturbations may push the freeze-out curve into unphysical region,  leading to unphysical results. On the other hand, the freeze-out curves obtained with AHYDRO, see Fig.~\ref{fig:r_AH_etabar3}, do not suffer from such problems, as the longitudinal pressure in the system smoothly approaches zero in this framework (being always positive).

An interesting result of our numerical analysis at large rapidities is that the BRSSS formulation, as compared to DNMR,  is closer to MIS. The difference between Eqs.~(\ref{pi_s_MIS}), (\ref{pi_s_DNMR}), and (\ref{pi_s_BRSSS}) resides in  the terms governed by $\tau_{\pi\pi}$ in (\ref{pi_s_DNMR}) and by $\lambda_1$ in (\ref{pi_s_BRSSS}). Both of them correspond to a shear-shear second-order coupling which is missing in the MIS approach. These terms are supposed to give a significant contribution far off equilibrium. In our case, although we start from a local equilibrium state, the rapid expansion of the system produces flow and pressure anisotropy very quickly and, especially at the edges of the system, both the shear flow tensor $\sigma^{\mu\nu}$ and the shear stress tensor $\pi^{\mu\nu}$ become significantly large.

We note that in the derivation of second-order hydrodynamic equations, the last term in (\ref{pi_s_BRSSS}), describing the 
$\pi^{\langle \mu \alpha}_{\vphantom{\alpha}} \pi^{\,\,\,\nu \rangle}_\alpha$ correction, is originally a 
$\sigma^{\langle \mu \alpha}_{\vphantom{\alpha}} \sigma^{\,\,\,\nu \rangle}_\alpha$ second-order term in the formula for $\pi^{\mu\nu}$, see Eq.~(3.11) in~\cite{Baier:2007ix}, while in the DNMR formulation the  $\sigma^{\langle \mu \alpha}_{\vphantom{\alpha}} \pi^{\,\,\,\nu \rangle}_\alpha$ term appears from the beginning as a direct consequence of using the kinetic theory. For systems being close to equilibrium, one may argue that the terms $\pi^{\langle \mu \alpha}_{\vphantom{\alpha}} \pi^{\,\,\,\nu \rangle}_\alpha$ and $\sigma^{\langle \mu \alpha}_{\vphantom{\alpha}} \pi^{\,\,\,\nu \rangle}_\alpha$, if multiplied by the appropriate kinetic coefficients, become equivalent (if one uses the first-order NS hydrodynamics approximation, $\pi^{\mu \alpha}_{\vphantom{\alpha}} = 2 \eta \sigma^{\mu \alpha}_{\vphantom{\alpha}}$). This has been recently confirmed by performing the gradient expansion for these hydrodynamics frameworks~\cite{Florkowski:2016zsi}. On the other hand, as our numerical calculations illustrate, in the situations far off equilibrium, the terms  $\pi^{\langle \mu \alpha}_{\vphantom{\alpha}} \pi^{\,\,\,\nu \rangle}_\alpha$ and $\sigma^{\langle \mu \alpha}_{\vphantom{\alpha}} \pi^{\,\,\,\nu \rangle}_\alpha$ cannot be simply exchanged with the help of the Navier-Stokes formula. This further indicates that the results of hydrodynamic equations may depend not only on the values of the kinetic coefficients but also on the very special choice of the terms included in the equations. A natural solution to this problem is the use of the most complete form of such equations.

\section{Conclusions}
\label{sect:conclusions}

In this work we have analyzed the non-boost-invariant hydrodynamic evolution of matter produced in heavy-ion collisions. The results obtained with several popular implementations of second-order dissipative hydrodynamics have been compared. In addition, a newly developed framework of anisotropic hydrodynamics has been also used. 

Our numerical calculations, assuming initial local equilibrium conditions and gaussian rapidity profiles,  indicate that all hydrodynamic approaches yield consistent results for the evolution of the central (mid rapidity) part of the created system. On the other hand, they vary substantially in their predictions for large rapidities. In this region, all standard formulations of relativistic hydrodynamics predict the appearance of a large and negative longitudinal pressure, which may lead to misleading conclusions concerning particle production. The largest negative pressures appear if one uses the MIS and BRSSS frameworks. The DNMR prescription works better, leading to smaller negative values of the pressure. Such differences may be connected with different treatment of the shear-shear coupling in various approaches. In all studied cases, a change from positive to negative pressure has the character of a shock, where pressure as well as the fluid rapidity change very suddenly in a narrow range of space-time rapidity $\eta$.  The regions with negative pressure may be also an obstacle for the determination of the freeze-out hypersurfaces, especially for event-by-event simulations and small systems. 

The approach of anisotropic hydrodynamics is free of this problem as all the pressure components are positive by construction in this framework.  As a result, anisotropic hydrodynamics may be used as a practical tool to regulate unphysical behavior at large rapidities. 

\medskip
\acknowledgments{
W.F. and L.T were supported by Polish National Science Center
Grant DEC-2012/06/A/ST2/00390.  R.R. was supported by Polish 
National Science Center Grant DEC-2012/07/D/ST2/02125.
M.S. was supported by the U.S. Department of Energy, Office of Science,
Office of Nuclear Physics under Award No. DE-SC0013470.
}

\bigskip
\appendix
\section{Third-order terms in the Israel-Stewart theory}
\label{sect:thirdorder}

In this Section we discuss Eq. (\ref{pi_s}) that, using Eq.~(\ref{en_mom_con_IS1}), may be cast in the form
\begin{eqnarray}
&& -\pi_s   \frac{T \beta_\pi}{2} \partial_\mu \left( \frac{1}{ T \beta_\pi} U^\mu\right) \nonumber \\
&=&  -\pi_s\frac{ \theta }{2}\left[1+ \left(\frac{{\cal E} +{\cal P} }{{\cal E}}  \right)\frac{{\cal E}}{T}\frac{  dT}{ d{\cal E}}\left( 1+\frac{T}{\beta_\pi}\frac{d\beta_\pi}{dT}\right)\right]\nonumber\\
&& +\pi_s\frac{ \theta }{2}\left(\frac{\pi_s}{{\cal E}}  \right)\frac{{\cal E}}{T}\frac{  dT}{ d{\cal E}}\left( 1+\frac{T}{\beta_\pi}\frac{d\beta_\pi}{dT}\right)
\label{lastterm1}
\end{eqnarray}
where in the last line we singled out the term proportional to $\pi_s^2 \theta$  which is of the third-order in gradients, see also~\cite{Heinz:2005bw,Shen:2014vra} . In this case we may use the equilibrium expressions ${\cal P}={\cal E}/3$ and ${\cal E}\propto T^4$. Moreover, for the RTA kinetic equation one obtains $\beta_\pi=4 {\cal P}/5$ and $\tau_\pi = 5 {\bar \eta}/T$. Hence, the right-hand side of  Eq.~(\ref{lastterm1}) can be rewritten as
\begin{eqnarray}
-\frac{4}{3} \pi_s\theta
+ \frac{5}{24}  \frac{\pi_s^2}{{\cal P}}    \theta.
\end{eqnarray}
This result used in Eq. (\ref{pi_s}) leads to the equation
\begin{eqnarray}
D\pi_s+\frac{\pi_s}{\tau_\pi}= \frac{16}{15}  {\cal P} \theta - \frac{4}{3} \pi_s    \theta +\frac{5}{24}  \frac{\pi_s^2}{{\cal P}}    \theta,
 \label{piseq}
\end{eqnarray}
which, after neglecting the last term, reproduces (\ref{pi_s_MIS}).

\bibliography{nbinv}

\end{document}